%% file: Main.tex
\documentclass[runningheads]{llncs}

\usepackage[hyphens]{url}

\usepackage{graphicx}
\usepackage{xcolor}
\usepackage{multirow}
\usepackage{tikz,hyperref}

\usepackage{amsmath,amssymb,amsfonts}

\usepackage{algorithmic}
\usepackage[linesnumbered,ruled,vlined]{algorithm2e}
\usepackage{lineno}

\usepackage{textcomp}

\usepackage[final]{changes}

\definechangesauthor[color=DarkCyan]{AMS}
\definechangesauthor[color=NavyBlue]{MB}
\definechangesauthor[color=Indigo]{JA}

\begin{document}

\title{End to End Secure Data Exchange in Value Chains with Dynamic Policy Updates}

\titlerunning{End to End Secure Data Exchange}

\author{Aintzane Mosteiro-Sanchez\inst{1,2} \and Marc Barcelo\inst{1} \and Jasone Astorga\inst{2} \and Aitor Urbieta\textbf{\inst{1}}}

\authorrunning{A. Mosteiro-Sanchez et al.}

\institute{Ikerlan Technology Research Centre, Basque Research and Technology Alliance (BRTA). P.º J.M. Arizmendiarrieta, 2. 20500 Arrasate/Mondragón, Spain \\
\email{\{amosteiro,mbarcelo,aurbieta\}@ikerlan.es} \and
Department of Communications Engineering, Faculty of Engineering, University of the Basque Country UPV/EHU,Alameda Urquijo s/n, 48013 Bilbao, Spain
\email{jasone.astorga@ehu.eus}}

\maketitle

\begin{abstract}
Data exchange among value chain partners provides them with a competitive advantage, but the risk of exposing sensitive data is ever-increasing. Information must be protected in storage and transmission to reduce this risk, so only the data producer and the final consumer can access or modify it. In most cases, data producers are IIoT devices, limited in terms of processing and memory capabilities. End-to-end (E2E) security mechanisms have to address this challenge, and protect companies from data breaches resulting from value chain attacks. Moreover, value chain particularities must also be considered. Multiple entities are involved in dynamic environments like these, both in data generation and consumption. Hence, a flexible generation of access policies is required to ensure that they can be updated whenever needed. This paper presents a Ciphertext-Policy Attribute Based Encryption (CP-ABE) reliant data exchange system for value chains with E2E security. It considers the most relevant security and industrial requirements for value chains. The proposed solution can protect data according to access policies and update those policies without breaking E2E security or overloading field devices. The experimental evaluation has shown the proposed solution's feasibility for IIoT platforms.
\end{abstract}



\input{Docs/1.Introduction}

\input{Docs/2.SoA}

\input{Docs/3.Scenario}
\input{Docs/4.ProposedSystem}
\input{Docs/5.PolicyUpdate}

\input{Docs/6.Experimental}
\input{Docs/7.Conclusions}

\input{Docs/ACKs}

\bibliographystyle{splncs04}

\bibliography{references}

\end{document}

%% file: Docs/1.Introduction.tex
\section{Introduction} \label{Sec:1}

\replaced{Value chains are the evolution of traditional supply chains. Value chains, in addition to physical assets, also manage digital assets and production parameters. The exchange of this information between retailers, customers, and manufacturers gives companies a competitive advantage, increases efficiency, and reduces production costs~\cite{Lu2017}. Integrating Industrial IoT (IIoT) devices into this infrastructure facilitates information management, but it also introduces new risks and vulnerabilities. For example, in 2016, over 200,000 industrial systems were exposed on Shodan~\cite{Andreeva2016IndustrialAvailability}. This search engine exposes IP addresses, open services, and vulnerabilities~\cite{9072175}. In value chains, this information can be used to compromise one of the partners. Due to the interconnected nature of value chains, the consequences of an attack on one member can spread to the rest~\cite{VIEIRA2020100161}, giving attackers access to the infrastructures of other stakeholders. Building a system that ensures secure data exchange between partners is a security challenge that requires meticulous design and planning.}{Value chains are the natural evolution of supply chains. In value chains, the exchange of goods between companies is not limited to physical assets but also involves digital assets and production parameters. Value chains provide companies with a competitive advantage, as exchanging information between retailers, customers or manufacturers is essential to increase efficiency and reduce costs in Industry 4.0~\cite{Lu2017}. As a result of the Industrial Internet of Things (IIoT) adoption in manufacturing environments, over two hundred thousand industrial systems have been exposed in Shodan~\cite{Andreeva2016IndustrialAvailability}. This search engine exposes IP addresses, open services, and vulnerabilities~\cite{9072175}. In a value chain, this information can be used to compromise one of the partners. Due to the interconnected nature of value chains, the consequences of an attack on one member can propagate to the rest~\cite{VIEIRA2020100161}, giving attackers access to the infrastructures of other participants. Building a secure value chain infrastructure between partners is thus a security challenge that requires meticulous design and planning. }

Value chain security is a growing concern for organizations reluctant to share sensitive data with partners who may be both collaborators and competitors. Protecting valuable information and controlling who has access to it are among the main security challenges for value chains~\cite{noauthor_what_2020}. \replaced{The 2022 Data Breach Report by IBM~\cite{IBM2022CostIBM} shows that 19\% of companies' data breaches are a result of supply chain attacks. Supply chain attacks harm companies' reputations and cause significant economic damage. In fact, data breaches resulting from a supply chain compromise are more expensive for companies. On average, a data breach costs USD 4.46 million, while a data breach resulting from a supply chain compromise costs USD 4.35 million, 2.5\% more than a standard data breach~\cite{IBM2022CostIBM}. Furthermore, the cost of a data breach, in general, has risen a 12.7\% since 2020~\cite{IBM2020CostIBM}.}{The 2020 Data Breach Report by IBM~\cite{IBM2020CostIBM} showed that attacks on value chain infrastructures can result in data breaches, harming companies' reputations and causing significant economic damage. The same report also indicated that the average estimated cost of data breaches for private companies in 2020 was €3.5 million.} These staggering numbers make value chain security of utmost importance.

\replaced{Ensuring the confidentiality of industrial data is essential to reduce the impact of data breaches. Protecting information during transmission and storage prevents attackers from obtaining sensitive information even if it is leaked. To do so, this protection must be end-to-end (E2E), i.e., only the generator and the receiver must have access to it.}{Protecting industrial data both in transmission and storage has a pivotal role in reducing the impact of data breaches.} Protocols like TLS or DTLS are the dominant strategy to secure data in transit. The major drawback of these protocols is that their security ends at every middlebox, allowing attackers access to the transmitted information through these network elements. \added{In addition, in environments where data created by one device can be consumed by multiple devices, TLS and DTLS present inefficiency issues. These environments benefit from one-to-many communication, and TLS/DTLS require encrypting information individually for each destination.}

Regarding stored data, some cloud service providers integrate cryptographic mechanisms based on encryption protocols such as AES or RSA. However, this can lead to vendor lock-in~\cite{Opara-Martins2016CriticalPerspective}, forcing every partner in the value chain to use the same service for information exchange to guarantee an end-to-end (E2E) secure exchange. 
\replaced{Therefore, the challenge to design a secure and efficient data exchange architecture for value chains capable of meeting E2E confidentiality remains. Moreover, the system must be flexible and adaptable to changes, as this responsiveness is considered essential for value chains~\cite{BICOCCHI2019111}, as well as guarantee data integrity, which is crucial to achieve accurate decision-making in supply chains~\cite{HAMALAINEN2019100105}.}{Therefore, it becomes imperative to have solutions that provide E2E security within a value chain and do not depend entirely on third-party services. To this end, Ciphertext-Policy Attribute-Based Encryption (CP-ABE)~\cite{waters2011ciphertext} has shown promising results.}

\replaced{With these goal in mind, one-to-many encryption schemes are of great interest to achieve the desired E2E confidentiality in distributed environments. In particular, non-identity-dependent schemes, such as Ciphertext-Policy Attribute-Based Encryption (CP-ABE)~\cite{waters2011ciphertext}, have shown promising results. CP-ABE is an encryption scheme that generates users' secret keys based on attributes and protects messages according to access policies defined with those attributes. This scheme allows multiple users to access the same encrypted message as long as their attributes comply with the policy, which increases efficiency. However, CP-ABE lacks the required flexibility since it does not natively integrate a policy update system. In response to this situation, our proposal has the following contributions:}{CP-ABE is an encryption scheme that protects messages according to policies and generates secret keys using attributes. The combination of attributes and policies allows multiple users to access the same encrypted message as long as their attributes comply with the policy. To optimize the potential of CP-ABE within a value chain, it is desirable to maximize its adaptability to system changes. This adaptability to change is defined as responsiveness and is considered essential in value chains~\cite{BICOCCHI2019111}. The access policy under which the original message was encrypted with CP-ABE must be updated to achieve this responsiveness. However, a well-known problem with CP-ABE is that the scheme itself does not consider this requirement. Our proposal applies a CP-ABE scheme that overcomes this limitation and updates policies without compromising E2E confidentiality and integrity. Data integrity is crucial to achieving accurate decision-making in supply chains~\cite{HAMALAINEN2019100105}. To achieve this, we use Multi-Layered CP-ABE~\cite{9460263}, based on the ETSI CCA Secure modular scheme~\cite{ETSI-CYBER2021AttributeV1.2.1}. Our main contributions are as follows.}
\begin{itemize}
    \item \replaced{We develop an efficient E2E secure information exchange system for value chains based on a one-to-many encryption scheme. Moreover, the proposal uses CP-ABE, but it is agnostic to a specific CP-ABE construction.}{We develop an E2E secure information exchange system for value chains agnostic to the underlying CP-ABE construction.}
    \item \replaced{We achieve the required flexibility by applying an access policy update scheme~\cite{9460263} that neither breaks E2E security nor regenerates the symmetric key, nor interacts with field devices.}{We apply an access policy update scheme that does not break E2E security, nor does it regenerate the symmetric key or interact with field devices.}
    \item \replaced{Our solution allows data producers to control who has access to their information without prior knowledge of user identities. Therefore, we favor scalability by allowing the addition of new users without generating additional operations to the IIoT devices that produced the data.}{Our solution gives data producers control over who has access to information. This control can be exercised without prior knowledge of the identity of each user. Our system, hence, favors scalability by allowing the addition of new users without involving the IIoT devices that produced the original data.}
    \item We prove the feasibility of our solution by deploying it on a resource-constrained device with ARMv6 architecture, emulating OT devices.
\end{itemize}

The rest of the paper is organized as follows: Section \ref{Sec:2} analyses the State of the Art. Section \ref{Sec:3} presents our scenario as well as the requirements defined for the system. Section \ref{Sec:4} discusses the proposed system and Section \ref{Sec:5} the used algorithms. We close the paper with an experimental evaluation in Section \ref{Sec:6} and conclusions in Section \ref{Sec:7}.

%% file: Docs/2.SoA.tex
\section{State of the Art} \label{Sec:2}



Security is one of the critical issues of value chain management. In particular, information exchange vulnerabilities are considered the primary security concern \replaced{for value chains}{in these scenarios}~\cite{doi:10.1080/13675567.2016.1219703}. \replaced{Real-world experiences show that such concerns are not unfounded: attacks to these infrastructures have been on the rise since 2020, exposing sensitive information and affecting critical industries~\cite{IBM2020CostIBM}.}{Real-world experiences show that such concerns are not unfounded: 2020 saw a surge in attacks on these infrastructures, exposing sensitive information and affecting vital industries~\cite{IBM2020CostIBM}.} Value chain attacks are expected to grow in frequency and severity in the following years, which leads to an increased volume of data breaches, exposed information, and affected users. This has been analyzed by the Identity Theft Resource Center (ITRC)~\cite{ITRC2021Q1Analysis}. According to their Data Breach Analysis report, during the first quarter of 2021, attacks on value chains grew by 42\% compared to the last quarter of the previous year. The attacks resulted in a rise of just 12\% in data breaches, but the number of impacted users jumped from 8 million to 51, increasing 564\%. As company interconnections grow, partners should upgrade their risk management strategies to include end-to-end security and protect themselves from compromised value chain members.

\replaced{Value chain security must protect physical assets, digital products, and associated production parameters.}{Value chain security has to evolve from protecting physical goods to securing digital products and associated production parameters.} Previous research~\cite{VAZQUEZMARTINEZ201890} combines cloud computing and a cryptographic envelope to protect product delivery. However, this approach is limited to transferring finished digital products and does not consider the exchange of manufacturing information. Researchers in~\cite{10.1145/3427228.3427248} combine Bloom filters and Oblivious Transfer to guarantee a private industrial parameters exchange. However, this solution reduces the control retained by data producers by allowing customers to retrieve data without the data producers knowing what has been transferred. Overall, there is a need for encryption schemes adaptable to Industrial IoT (IIoT) devices and capable of providing E2E security~\cite{Mosteiro-Sanchez2020Securing4.0}. This requirement is magnified in value chains dealing with industrial limitations and risks introduced by third-party services.

\replaced{Managing data security by ensuring data confidentiality reduces the exposure of sensitive information and limits the impact of data breaches.}{Addressing data security can reduce sensitive information exposure and limit the impact of data breaches.} However, industrial data encryption is a sensitive issue in manufacturing due to the large volume of data to be managed and exchanged~\cite{Liu2011LeveragingChains}. To address this issue, authors in~\cite{8972565} combine Blockchain and symmetric encryption. The proposed system \replaced{achieves the mentioned one-to-many encryption,}{allows authors to encrypt for multiple receivers,} but the identity of \replaced{the receivers must be known in advance.}{those receivers must be known from the beginning.} This limitation hinders scalability, making it difficult to add new users, and can have an unpredictable impact on system efficiency.

\replaced{Regarding logistics security, the authors of~\cite{9162337} have developed a secure E2E sensing system for supply chains. Their solution focuses on ensuring that sensor readings are reliable and that each parameter is related to an existing physical event.}{Regarding logistics security, authors in~\cite{9162337} have developed a secure E2E sensing system for supply chains. Their solution gives particular relevance to whether the sensor readings are trustworthy and every saved parameter is directly related to a physical event in the real world.} Because the proposal is limited to sensor readings and does not consider other types of data, its direct application to digital assets in a value chain is not straightforward\deleted{, reducing the company's competitive advantage}. Looking for E2E security, researchers in~\cite{10.1145/3445969.3450423} develop a security system for publish/subscriber communications in cyber-physical systems. However, this proposal is designed for the Operational Technology (OT) network, not a value chain. 

\replaced{In general, the different security proposals introduce interesting considerations. For example, the data exchange solution must ensure that data producers can determine who can access the information. Moreover, they must do so without knowing each recipient's identity in advance. Similarly, the solution must be scalable, consider IIoT devices, and must not assume that every data exchange is tied to an event in the physical world.} {Overall, the different security proposals introduce interesting concerns, some of which should be considered: For example, having data access control provides the system with greater flexibility for one-to-many communications.}

\replaced{Given the points mentioned in the previous paragraph, CP-ABE offers an optimal solution. It offers a one-to-many encryption system that provides the sought-after E2E confidentiality and allows Data Owners (DO)s to maintain control of their data. This control is achieved by protecting information with attribute-based access policies without binding decryption to a specific data requester. In addition, this system allows new users to access old information, promoting scalability.}{To this end, CP-ABE is an excellent option to achieve E2E confidentiality as well as more adaptable access control for one-to-many encryption. Moreover, combined with symmetric encryption like AES-GCM, it also provides E2E integrity protection. For these reasons, in this paper, we focus on CP-ABE. See Section \ref{Sec:3} for further analysis and details. CP-ABE can provide the sought E2E confidentiality and allow DOs to retain data control. This control is achieved by protecting information with access policies based on attributes, without linking decryption to a specific data requester. Besides, this system allows new users to access old information, which favors scalability.} 

\deleted{With this in mind, }Authors in~\cite{7442163} combine CP-ABE with a symmetric cipher to incorporate encryption to attribute-based access control. However, they focus on protecting the RFID tags attached to the products. The requirement of RFID tags makes it difficult to apply it to our scenario, which includes\deleted{both} data and digital goods. Authors in~\cite{9145100} combine CP-ABE with Blockchain in international supply chains. They consider a product exchanged through different partners, but to which not all of them can access. \replaced{Overall, applying CP-ABE to a trade infrastructure provides E2E security and allows one-to-many encryption. However, to guarantee the responsiveness and flexibility mentioned in the introduction, the policies used in CP-ABE should be updatable while preserving E2E security.}{In general, the application of CP-ABE to a trade infrastructure must ensure E2E security and allow one-to-many encryption. Moreover, to have long-lasting encryption at rest, the access policies applied to encrypted data should be updatable while preserving E2E security.}

The \replaced{difficulty of updating}{update of} access policies is a known issue in CP-ABE. One of the works identifying it is~\cite{10.1007/978-3-319-44482-6_1}, where authors solve it with a layered model allowing policy updates. However, their model requires knowing the information recipients beforehand. Another approach is~\cite{Yasumura2018}, in which the authors focus on re-encrypting the symmetric ciphertext while DOs must produce a new ABE ciphertext with every update. However, the computational burden placed on DOs for updates makes this solution inadequate for IIoT devices. \deleted{Researchers in {Al-Dahhan2019SurveyEncryption} consider the Cloud an untrusted environment and require having control over the data, although they do not discuss the need to update access policies.} \replaced{Authors in~\cite{Li2019AnComputing}}{An approach that considers access policy updates is~\cite{Li2019AnComputing}, in which authors} modify the linear secret sharing scheme used to define the access policy embedded in the encrypted message. This update, however, has to be performed by DOs, which can have an unpredictable computational cost for them. To reduce the burden on DOs,~\cite{8766597} proposes a hybrid system in which the ciphertext is sliced, and one of the slices is randomly chosen to be updated. Thus, nothing stops an attacker from collecting slices at different stages and then combining and decrypting them. Finally, authors in~\cite{cmc.2020.06278} combine CP-ABE with symmetric encryption. \replaced{However, they use the same symmetric key for every message. Therefore, when the system updates the access policy, authors deem it necessary to update the symmetric key. This forces the DO to regenerate the symmetric key and re-encrypt it with CP-ABE.}{However, they apply the same symmetric key to every message and thus consider that the symmetric key has to be updated whenever an access policy changes. Thus, they force the DO to generate a new symmetric key and encrypt it with CP-ABE again.}

Therefore, there appears to be no widely efficient data exchange method for value chains that support non-identity-based access to data with an efficient policy update mechanism. \replaced{To address this, we set up a CP-ABE-based E2E secure data exchange that allows DOs to control who accesses their data without identifying users in advance.}{In order to solve this, we establish an E2E secure data exchange that allows DOs to retain control over who access data without having to identify the users individually.} Since the information recipients do not have to be known beforehand, we favor scalability by allowing easy addition of users to the system. This also reduces the burden placed on DOs, who do not need to negotiate keys with every user in the system nor reencrypt ciphertext when new users are added. Finally, the policy update provides the system with resilience without needing intervention from DOs.

%% file: Docs/3.Scenario.tex
\section{Scenario and Requirements}\label{Sec:3}

\begin{figure}[htbp]
\centerline{\includegraphics[width=85mm]{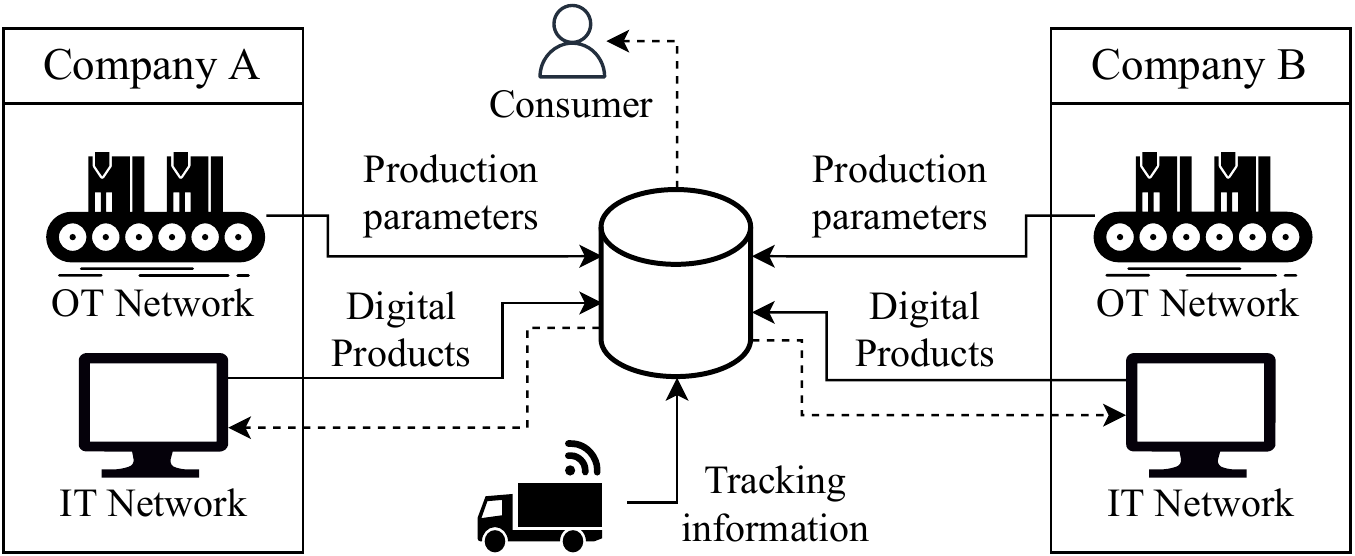}}
\caption{Value Chain data exchange situation layout}
\label{fig:FIG1}
\end{figure}

\deleted{The term ``value chain" refers to all the steps required to create a physical or digital product. In other words, it extends beyond the physical product exchange and also encompasses all the information generated during product design and development. Therefore, the use case presented in this paper covers the exchange of digital and physical assets and the parameters generated during their creation.} \replaced{Value chains are complex distributed systems bounded by industrial requirements. Therefore, the companies participating in a value chain play roles associated with distributed information exchange systems. Thus, we can speak of information consumers, generators, and consumers-generators. Figure \ref{fig:FIG1} presents a schematic of a generic value chain with companies, consumers, and transport companies.}{Figure \ref{fig:FIG1} illustrates a generic value chain in which multiple entities are involved. In it, we identify three main roles for our use case: companies, transport or storage systems, and consumers. Companies create and process physical and digital products. Transportation and storage systems exchange and store physical products between different value chain members, generating data associated with transportation and storage conditions. Finally, data consumers do not produce information for the value chain but only consume the information generated. This data consumption is represented by a dotted line in Figure \ref{fig:FIG1}, while the solid line represents data generation and storage.}

\added{As mentioned above, the different participants in the value chains fulfill existing roles in distributed information exchange systems. Thus, storage and transport companies are generators: they produce and store data related to the product's transport and storage conditions. End users and after-sales services constitute pure information consumers. And finally, companies are consumer-generators. In Figure \ref{fig:FIG1}, information consumption is represented by a dotted line and information generation by a solid line.}

\replaced{Companies generate two data types: manufacturing parameters and digital products. Digital products are developed with high-performance equipment, but manufacturing parameters are generated by IIoT devices. Thus, the selected solution must be scalable to IIoT devices.}{Companies generate two data types: manufacturing parameters and digital products. Manufacturing parameters related to physical products are generated in the OT network by IIoT devices. As for digital products, they are generated by high-performance equipment in the IT network. In both cases, the integrity of the information must be guaranteed, as well as E2E confidentiality. Thus, the solution must be scalable to IIoT devices.} Companies protect and store the information in a storage solution accessible to other chain members. In addition to being producers, companies are also consumers: they access digital products and production parameters generated by other companies. The data exchange platform must ensure that companies only have access to data they are authorized to obtain.

Regarding transportation companies, they are responsible for trading physical products. Physical product exchange and storage generates data of significant interest for the remaining members of the value chain. Examples of this are product geolocation, as well as parameters concerning transport and storage conditions, such as temperature or humidity. This role is limited to information generation and storage and does not consume shared data.

\subsection{Design Requirements}\label{Sec:3.1}

\replaced{The main goal is to develop a secure E2E data exchange for a value chain. To this end, this section defines the five requirements to be met by the developed solution. These requirements have been established taking into account the industrial constraints of value chains, as well as the needs identified in section \ref{Sec:2}. Table \ref{Tab1:Requirements} analyzes, to the best of our knowledge, which requirements are met by the current state-of-the-art.}{The primary goal is to develop an E2E secure data exchange for a value chain. This section defines five requirements to achieve this goal, considering both security and industrial limitations. Table \ref{Tab1:Requirements} analyzes, to the best of our knowledge, which requirements are met by cases studied in Section \ref{Sec:2}.}

\begin{table}[ht]
\renewcommand{\arraystretch}{1.2}
\caption{Defined requirement fulfillment level.}
\label{Tab1:Requirements}
\centering
\resizebox{89mm}{!}{%
\begin{tabular}{c|ccc|cccc}
\hline 
\multirow{2}{*}{Approach}                               & \multicolumn{3}{c|}{R1}                               & \multirow{2}{*}{R2}   & \multirow{2}{*}{R3}   & \multirow{2}{*}{R4}   & \multirow{2}{*}{R5}
\\ \cline{2-4} 
                                                        & R1.1             & R1.2           & R1.3              &                       &                       &
\\ \hline \hline
                                            
Vazquez-Martinez et al.~\cite{VAZQUEZMARTINEZ201890}    & $\square$        & $\blacksquare$ & $\square$         & $\square$             & $\blacksquare$        & $\blacksquare$        & $\square$             \\ \hline
Pennekamp et al.~\cite{10.1145/3427228.3427248}         & $\blacksquare$   & $\square$      & $\square$         & $\blacksquare$        & $\square$             & $\blacksquare$        & $\square$             \\ \hline
Epiphaniou et al.~\cite{8972565}                        & $\blacksquare$   & $\blacksquare$ & $\blacksquare$    & $\square$             & $\blacksquare$        & $\square$              & $\square$             \\ \hline
Pennekamp et al.~\cite{9162337}                         & $\square$        & $\square$      & $\blacksquare$    & $\blacksquare$        & $\blacksquare$        & $\square$             & $\square$             \\ \hline
Dahlmanns et al.~\cite{10.1145/3445969.3450423}         & $\blacksquare$   & $\square$      & $\square$         & $\blacksquare$        & $\blacksquare$        & $\square$             & $\square$             \\ \hline
Qi et al.~\cite{7442163}                                & $\square$        & $\square$      & $\blacksquare$    & $\blacksquare$        & $\blacksquare$        & $\blacksquare$        & $\blacksquare$        \\ \hline
Pennekamp et al.~\cite{9145100}                         & $\square$        & $\square$      & $\blacksquare$    & $\square$             & $\blacksquare$        & $\blacksquare$        & $\square$             \\ \hline
\end{tabular}
}
\end{table}

\replaced{\textbf{R1. Heterogeneous Data Exchange.}}{\textbf{R1. Product and data exchange.}} \replaced{We identify three information exchange cases, and consider that the final solution should be extensible to any of them. The cases are listed below.}{We identify three instances of information exchange. The final solution should be extensible to any of them. They are listed below.}
\begin{description}
    \item \textbf{R1.1 Production parameters exchange.} This information exchange provides members of value chains with an advantage over their competitors. The system has to guarantee that sensitive data is managed securely.
    \item \textbf{R1.2 Digital product exchange.} Digital products must be securely transferred to the final consumer, assuring confidentiality and integrity.
    \item \textbf{R1.3 Physical product exchange.} The traditional exchange, linking the value chain with the conventional supply chain. The trade of physical products generates a multitude of sensitive data, which, if not protected, could be used by attackers and competitors.
\end{description}

\replaced{\textbf{R2. Computational Scalability}}{\textbf{R2. Working with limited computational power.}} \replaced{Data might be}{In the worst-case scenario, data is} generated by IIoT devices. These devices have a limited computational capacity, and the solution must be scalable \replaced{to achieve the required security operations.}{so that the required security operations are achievable.}

\textbf{R3. E2E data confidentiality and integrity.} \replaced{E2E security means that information must be secured before leaving the device that generated it. Therefore, data must be encrypted at the source by the device that generated it and maintain that encryption during transmission and storage.}{The information must be secured since it leaves the device that generated it. For this reason, data must be encrypted at the source and protected in transmission and storage. }

\textbf{R4. Non-Identity Based Access \deleted{Control }to Data.} Access to sensitive information must be controlled. However, \added{Section \ref{Sec:2} showed that} identity-based access control generates scalability and management issues in value chains. Thus, environments with many participants favor more scalable and flexible \replaced{approaches}{systems}. To this end, Role-Based Access Control (RBAC) or Attribute-Based Access Control (ABAC) provide fine-grained access to data and are easier to manage. Both support the definition of access policies unrelated to individual identities and instead use users' roles in the system or attributes associated with them.

\replaced{\textbf{R5. System flexibility and responsiveness.}}{\textbf{R5. Access Policy Update.}} \replaced{The data protection system must be flexible and capable of evolving when the access policies change.}{Access Control has to be flexible to define access policies and manage them.} Thus, to maximize the lifespan of the defined \deleted{access control }system, a policy update system should be in place. This system should consider both the case of a required policy update and any security event: from key renovations to security violations.


\replaced{Achieving E2E security in a supply chain involves meeting various requirements. However, as shown in Table \ref{Tab1:Requirements}, none of the approaches studied in Section \ref{Sec:2} can meet them all. Secure data exchange solutions must balance security, industrial availability, and system flexibility.}{Achieving E2E security in a supply chain involves considering industrial, security, and communication requirements. As shown in Table \ref{Tab1:Requirements}, none of the approaches studied in Section \ref{Sec:2} can meet all requirements.} Secure data sharing solutions must balance security, industrial availability, and system flexibility. 

%% file: Docs/4.ProposedSystem.tex
\section{Proposed System}\label{Sec:4}
\replaced{We define the proposed solution}{The proposed system has to be defined} considering the requirements defined in Section \ref{Sec:3.1}. To this end, this Section explains the encryption cipher choice, defines the roles of the proposed system, and how they are mapped to an industrial high-level reference model.

\subsection{Multi-Layered CP-ABE} \label{Sec:4.1}
\replaced{As analyzed in Section \ref{Sec:2}, CP-ABE is a promising one-to-many encryption scheme }{As explained in the Introduction, CP-ABE is an encryption scheme }that protects data by applying access policies—e.g., \textit{AP=(Mechanic AND Staff)}. It also generates users' secret keys using attributes—e.g., \textit{(Mechanic $||$ Staff $||$ Boss)}. Thus, the application of CP-ABE fulfills \textbf{R4} by creating an Attribute-Based Access Control to Data. However, CP-ABE by itself only provides confidentiality. Thus, it has to be combined with symmetric ciphers that also provide integrity (\textbf{R3}). In this paper, the chosen symmetric cipher is AES-GCM.

\replaced{Section \ref{Sec:3} established that the data exchange system must be flexible and capable of handling a policy update. However, this requires managing existing encrypted data without breaking the E2E confidentiality or integrity.}{Related to having an access control system is the requirement for the policies defined in the system to be updatable \textbf{R5}. Having access policy updates in CP-ABE implies dealing with already-encrypted data.} Thus, \replaced{the proposed solution requires}{we have to apply} a CP-ABE system that accounts for this and manages policy updates on \replaced{CP-ABE ciphertexts ($CT_{ABE}$)}{ciphertexts (CTs)} without overloading IIoT devices. To achieve this, we apply Multi-Layered CP-ABE~\cite{9460263}, which combines CP-ABE with a symmetric cipher and policy update without violating the original message's confidentiality or integrity. \deleted{Multi-Layered CP-ABE is based on ETSI's secure hybrid CCA encryption/decryption for CP-ABE~\cite{ETSI-CYBER2021AttributeV1.2.1}.}

\begin{figure}[!htbp]
\centerline{\includegraphics[width=55mm]{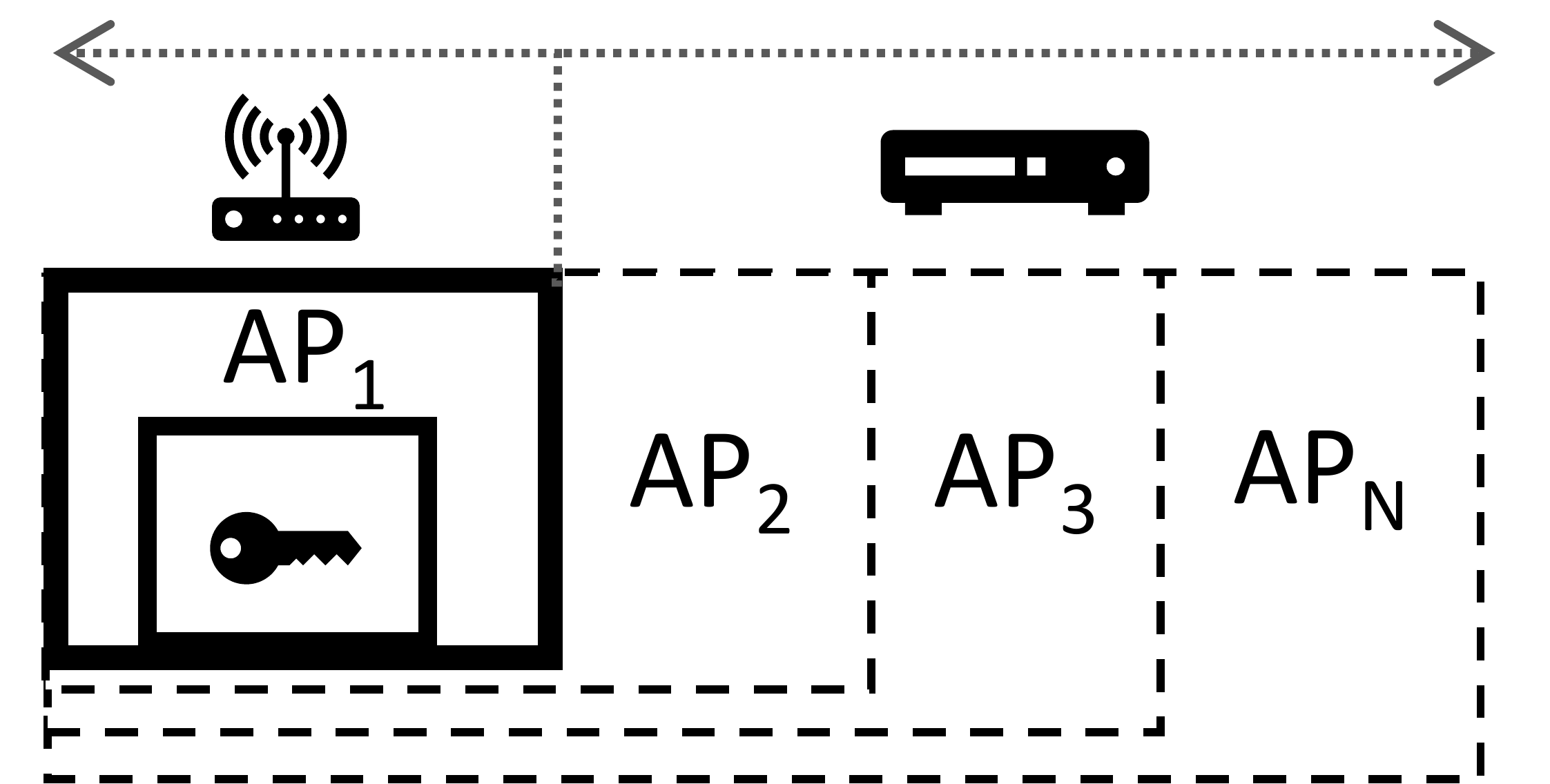}}
\caption{Layered encryption. The IIoT devices applies $AP_{1}$. Any other device can apply the following layers.}
\label{fig:FIG2}
\end{figure}

\replaced{Multi-Layered CP-ABE achieves policy updates through a layered encryption system, schematized in Figure \ref{fig:FIG2}. The original message ($PT$) is encrypted using symmetric encryption, e.g., AES. Then, the symmetric key ($SK_{sym}$) is encrypted using Multi-Layered CP-ABE. This implies iterative encryption of $SK_{sym}$. The first layer ($AP_{1}$) is immutable and has CCA security, which protects the $SK_{sym}$ against passive and active attackers. Then, the following layers ($AP_{2}$ through $AP_{N}$) are added, containing the policies that can be updated.}{Multi-Layered CP-ABE achieves policy update through a layered encryption system, Figure \ref{fig:FIG2}. The message is encrypted iteratively, and each encryption layer contains a different access policy. The outer layers contain the most dynamic policies ($AP_{N}$), and the inner layers contain the most stable ones—e.g., $AP_{3}$ is more variable than $AP_{2}$ but less variable than $AP_{N}$. This way, the most dynamic policies can be updated without regenerating the still valid ones. In order to reduce the interaction with IIoT devices, this encryption scheme establishes a first encryption layer that cannot be updated and has CCA security, which protects it against active and passive attackers.}

\added{The multilayer system requires a minimum of two layers, although developers can implement as many as desired. Users should note that fewer layers do not always lead to faster policy updates. The choice of the number of layers will depend on the device performing the update, the complexity of the policies contained in the layers, or the frequency with which they will be updated. Therefore, as suggested in~\cite{9460263}, the outermost layers should be defined according to policies with the highest variability.}

The layered system means that \replaced{$CT_{ABE}$}{the CP-ABE ciphertext ($CT_{ABE}$)} can be modified when an access policy is updated without exposing \replaced{$SK_{sym}$}{the symmetric key.} This system ensures the confidentiality and integrity of the original message. Finally, Multi-Layered CP-ABE reduces the computational burden on devices with more limited capabilities. This is because although the initial layer must be applied by the original device to guarantee E2E confidentiality, subsequent layers can be added by a more powerful device. These last layers can be updated and revoked as required by the system. 

\subsection{System Design} \label{Sec:4.2}
\deleted{The proposed system applies Multi-Layered CP-ABE~\cite{9460263}, based on ETSI's secure hybrid CCA encryption/decryption~\cite{ETSI-CYBER2021AttributeV1.2.1}.} Multi-Layered CP-ABE provides both attribute-based access to data (\textbf{R4}) as well as policy update \added{(\textbf{R5}).} Figure \ref{fig:FIG3} shows a conceptual representation of the E2E secure data exchange system. The developed system consists of eight roles, defined as follows: 

\begin{figure}[!htbp]
\centerline{\includegraphics[width=90mm]{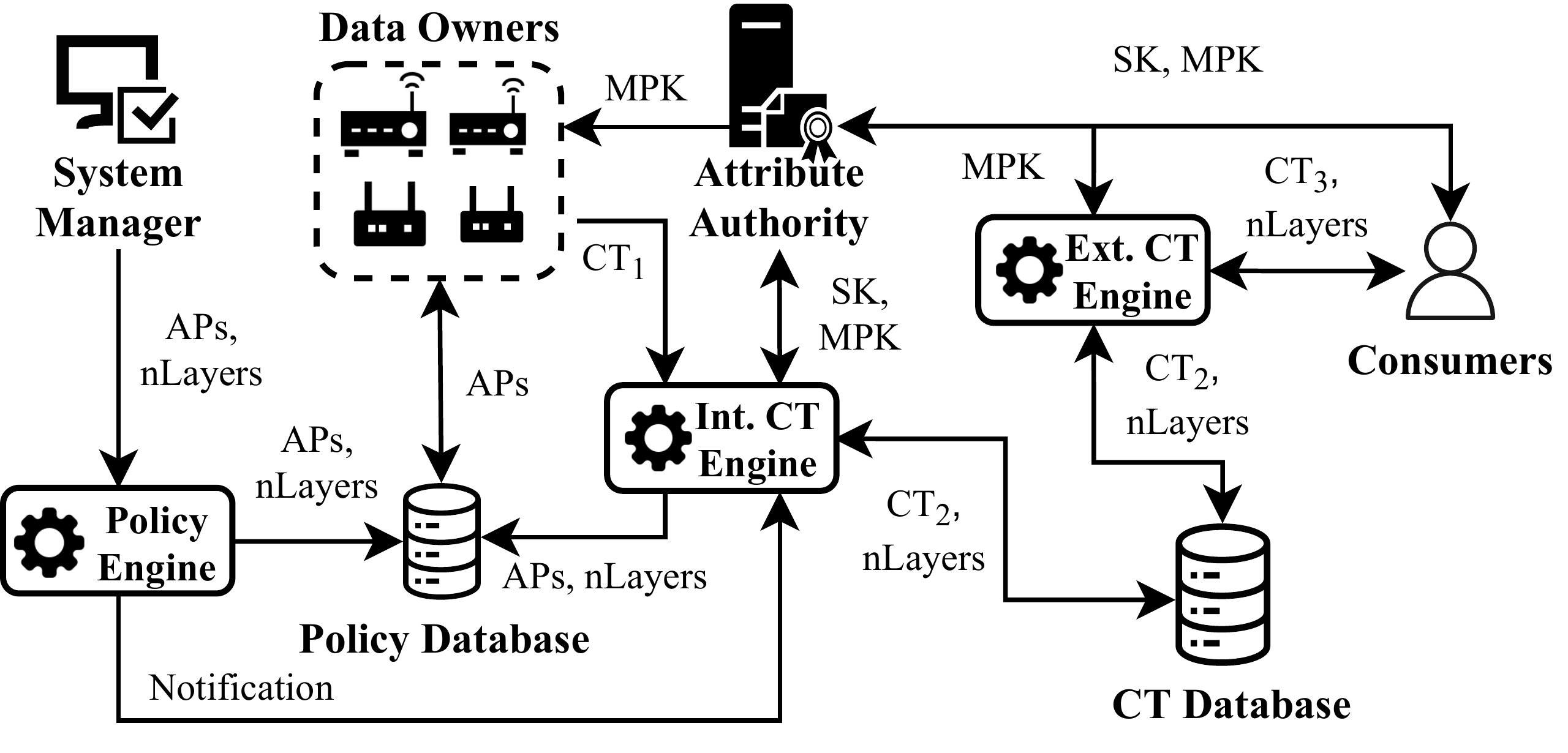}}
\caption{System Design. AP stands for access policy, CT for ciphertext, and nLayers is the number of access policy layers.}
\label{fig:FIG3}
\end{figure}

\begin{itemize}
    \item \textbf{Data Owners (DO)s.} \replaced{These are typically IIoT devices that generate the original data. They use a symmetric algorithm, e.g., AES, to protect it. Then, they use CP-ABE to encrypt $SK_{sym}$. This way, only the original device, and the intended consumers know $SK_{sym}$ (\textbf{R3}).}{They generate and encrypt data with a symmetric algorithm, e.g., AES. Only the original device and the intended consumers know the symmetric key (\textbf{R3}).} Data can be production parameters, digital products, tracking information, or any other sensitive information (\textbf{R1}). \replaced{$SK_{sym}$}{The symmetric key} is protected by an immutable policy, \deleted{with a first encryption layer, }as explained in Section \ref{Sec:4.1}. DOs do not intervene again, reducing the interaction with the capacity-limited devices (\textbf{R2}).
    \item \textbf{Attribute Authority (AA).} During system setup, the AA generates the Master Secret Key (MSK) and Master Public Key (MPK)\added{ for CP-ABE}. It stores and protects the MSK and sends the MPK to the DOs and CT engines. Finally, it generates the CP-ABE secret keys ($SK$s) based on consumers' attributes and the timestamp indicating the key's generation time. \deleted{Thus, each time a security event is recorded in the system, the consumer needs to renew his SK.}
    \item \textbf{System Manager.} \replaced{It manages system access policies. When a policy is updated, it is in charge of sending it to the policy engine.}{It manages the system access policies, defined according to the roles identified for the value chain. When a policy update takes place, it sends the new policy to the policy engine.}
    \item \textbf{Policy Engine.} It pushes the new policies to the Policy Database. It also notifies the Internal CT Engine when a policy update occurs.
    \item \textbf{Internal CT engine.} It adds the encryption layers containing all the revocable and updatable policies. Once encrypted, it pushes them to the CT database. Whenever an access policy update occurs, it receives a notification from the Policy Engine and retrieves the old ciphertext from the CT database, updating and storing it again (\textbf{R5}). \added{Note that} the DO does not need to intervene in the policy update process (\textbf{R2}).
    \item \textbf{External CT engine.} \replaced{Once system users obtain their CP-ABE key reflecting their attributes, they can use it to access all information whose access policy they comply with. However, there is always the risk of the keys being compromised. Therefore, it is necessary to have a system in place to ensure that consumers with old or compromised keys do not have access to the information. To this end, when a consumer requests a piece of data, the External CT Engine adds a new layer of CP-ABE encryption before sending it to the requesting consumer. This new layer defines an access policy that requires users to possess a key generated after the last security event in the system.}{To ensure that consumers with old keys do not have access to the information, it adds a new encryption layer. This new layer is defined according to a timestamp indicating the last security event recorded in the system. This way, only consumers with keys generated after the event can access the information, even if they comply with the access policies defined in inner layers.}
    \item \textbf{Policy Database.} It stores the access policies to be used for encryption.
    \item \textbf{CT database.} It stores the ciphertexts. \added{The ideal solution is to set up a distributed storage solution. This way, the CTs are always accessible, even if one of the nodes goes down. Solutions like IPFS may be suitable since it distributes the storage, is immutable, and is tamper-resistant.} 
\end{itemize}

\subsection{Industrial High-Level Reference Model} \label{Sec:4.3}

This section maps the roles shown in Figure \ref{fig:FIG3} to a high-level industrial reference model. \added{Industrial environments such as value chains benefit from defining the relationship between the architecture proposed in section \ref{Sec:3} and existing reference models. The definition of the relationship between architecture and model reduces the proposal's complexity and helps the stakeholders in the value chain define the structure to be used~\cite{Megow2020}.} \replaced{Thus, this mapping}{This mapping} allows the different companies in the value chain to coordinate their elements and functions, identify common elements and share them. \deleted{Moreover, it identifies the elements that every company in the supply chain must have.}

Regarding high-level reference models, ENISA adapted the Purdue Model to smart manufacturing by establishing a five-layer model~\cite{EuropeanUnionAgencyforNetworkandInformationSecurityENISA2018GoodManufacturing}. In this model, level 0 stands for the manufacturing process, levels 1-2 for the OT network, 3 for the OT-IT connection, 4 for the IT network, and 5 for third-party services.

\begin{figure}[htbp]
\centerline{\includegraphics[width=90mm]{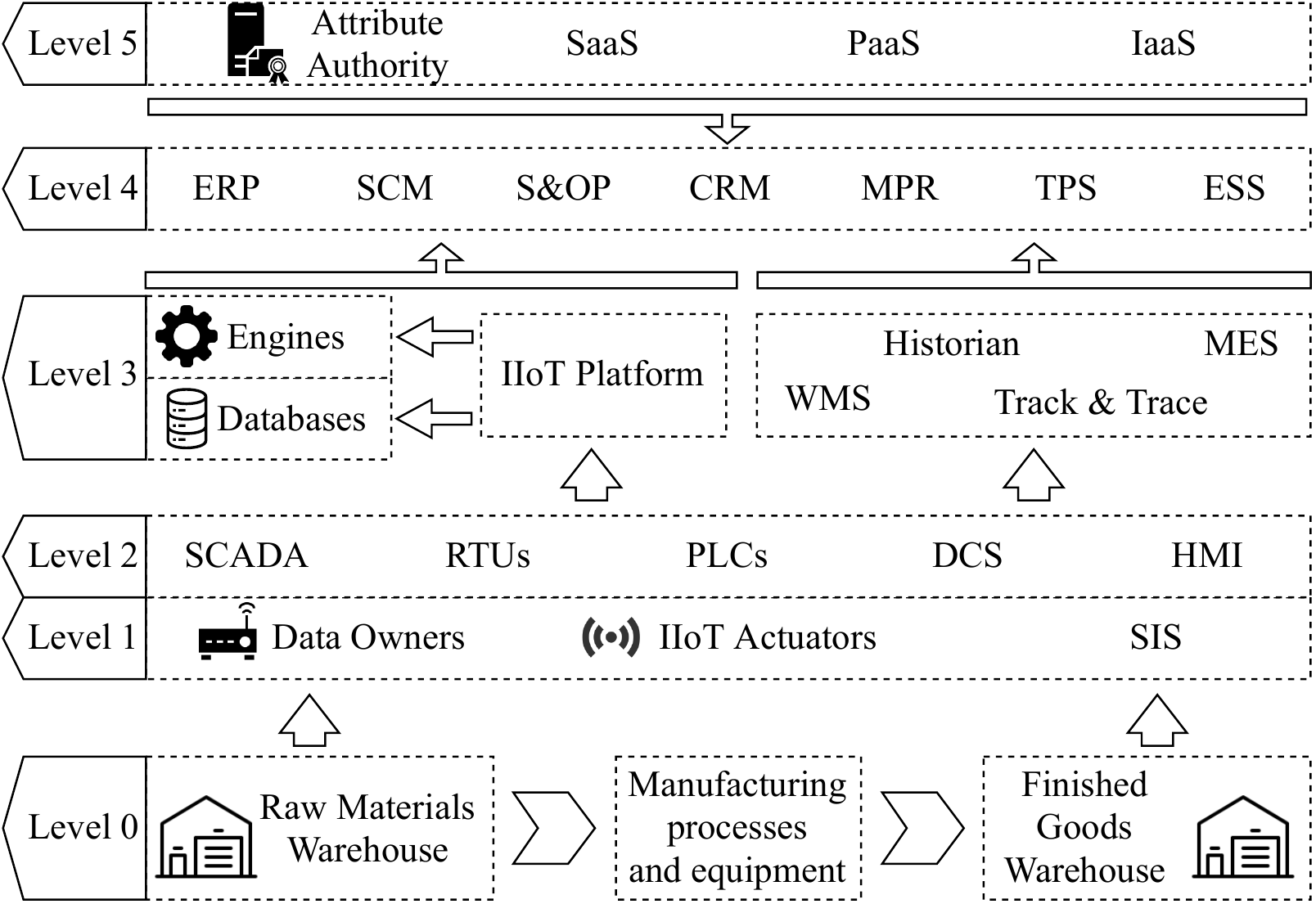}}
\caption{Components of the proposed system mapped to the ENISA high reference model~\cite{EuropeanUnionAgencyforNetworkandInformationSecurityENISA2018GoodManufacturing}.}
\label{fig:FIG4}
\end{figure}

Based on the ENISA model, Figure \ref{fig:FIG4} illustrates how the components of the proposed system are mapped to the high-level model. Data is generated and encrypted by the DOs, which can be IIoT devices located at level 1. The encrypted information is sent to the IIoT platform that the ENISA model defines at level 3. Finally, data reaches the engines and databases. The last element \replaced{to be mapped in the model}{to map to the model} is the AA. The encryption system is based on the MPK and the MSK used by the authority to generate individual secret keys. \replaced{The system presented has a single AA, shared by all members of the value chain and implemented through third-party services. In this context, the authority should be placed at level 5 of the ENISA model.}{The presented system is agnostic to the authority's implementation mode since it only requires it to perform the assigned functions. We consider a single authority shared by all the value-chain members and implemented through third-party services. In this context, the authority should be placed at level 5 of the ENISA model.}



%% file: Docs/5.PolicyUpdate.tex
\section{Policy Update and Revocation using Multi-Layered CP-ABE}\label{Sec:5}

The proposed CP-ABE scheme consists of \replaced{four}{five} algorithms: system setup, key generation, encryption, and decryption. Policy addition and policy revocation are achieved by encryption and decryption algorithms. 

\begin{itemize}
    \item $System Setup (K) \rightarrow MPK, MSK$ \par
    \replaced{This}{It} is the original \added{CP-ABE} setup algorithm, performed by the AA. After obtaining $MPK$ and $MSK$, it sends the $MPK$ to the DOs and the Engines. 
    
    \item $Key Generation (MSK, \mathbb{A}) \rightarrow SK$\par 
    \added{Whenever users request a CP-ABE Secret Key ($SK$), the AA generates a timestamp for the request ($T_{SK}$).} Then, the AA generates $SK$ using the $MSK$, according to an attribute set \added{$\mathbb{A}$ that contains the user's attributes and $T_{SK}$:} $ \mathbb{A} \gets (Att_{1}||Att_{2}||...||Att_{n}||T_{SK})$. \replaced{The inclusion of $T_{SK}$ in the policy}{This $T_{SK}$} makes sure that consumers with old $SKs$ cannot access the system.
    
    \item $Encryption (MPK, AP, PT) \rightarrow CT_{2}$\par
        \begin{figure}[!htbp]
        \centerline{\includegraphics[width=70mm]{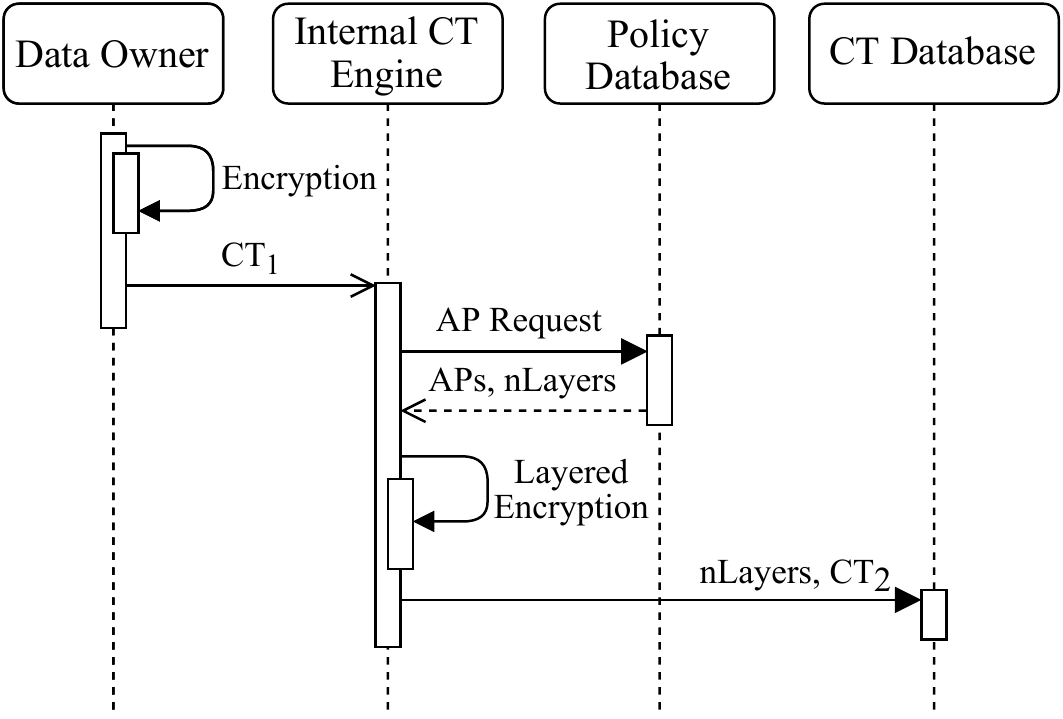}}
        \caption{Encryption message exchange}
        \label{fig:FIG5}
        \end{figure}
    
        \replaced{The message exchange between the DO and the internal CT engine during encryption is shown in Figure \ref{fig:FIG5}. Note that generally, the DO is an IIoT device. The encryption process takes two significant steps. The first one generates $CT_{1}$ and is explained below.}{The message exchange between the involved parties during encryption can be seen in Figure \ref{fig:FIG5}. The DO performs the original ETSI CCA Secure modular encryption~\cite{ETSI-CYBER2021AttributeV1.2.1} mentioned in the Introduction.} 
        \begin{enumerate}
            \item \replaced{The DO}{It} chooses a random AES-GCM key $SK_{sym} \in \{0,1\}^{n}$ and nonce $r \in \{0,1\}^{n}$. \added{The DO generates a new $SK_{sym}$ for every piece of data they create and encrypt.}
            \item \added{The same device} applies the Fujisaki-Okamoto transformation~\cite{Fujisaki2000} combined with the Access Policy $AP$ to create the nonce $u \gets H(r||SK_{sym}||AP)$. 
            \item \replaced{To complete the Fujisaki-Okamoto transformation, the DO appends $SK_{sym}$ to $r$, resulting in $M \gets (SK_{sym}||r)$. Then, the first layer of CP-ABE encryption is applied to $M$. This process of encrypting $SK_{sym}$ is also known as encapsulation. This encryption process is summarized as follows:}{The symmetric key is encrypted using CP-ABE:} $CT_{ABE_{1}}$ $\gets$ $Enc_{ABE_{CP}}$ $(MPK, AP, M, u)$. \added{The performed Fujisaki-Okamoto transformation ensures the CCA Security, which, as mentioned in Section \ref{Sec:4.1}, protects $SK_{sym}$ against passive and active attackers.}
            \item \added{After encapsulating $SK_{sym}$,} the original message, $PT$, is encrypted using AES-GCM\deleted{ to protect confidentiality and integrity}. \replaced{To guarantee the integrity of the encrypted $SK_{sym}$,}{To this end,} the authentication data is defined as $AAD$  $\gets$ $ExtractHeader$ $(CT_{ABE_{1}})$. \replaced{$PT$}{The message} is encrypted by running $CT_{AES}$ $\gets$ $Enc_{AES}$ $(SK_{sym},$ $PT, AAD)$. 
            \item Finally, the devices send the resulting ciphertext $CT_{1}$ $\gets$ $(CT_{AES},$ $CT_{ABE_{1}})$ to the Internal CT Engine. \added{With this, we finish the first exchange in Figure \ref{fig:FIG5}.}
        \end{enumerate}

        \added{The second step of the encryption process generates $CT_{2}$.} The internal CT engine adds the policy layers, which are applied so the innermost ones are less variable than the outermost ones. These layers contain the access policies that can be updated or revoked. \deleted{The engine performs the following computations:} \par
        
        \begin{algorithm} [!ht]
        \newcommand\mycommfont[1]{\footnotesize\ttfamily\textcolor{blue}{#1}}
        \SetCommentSty{mycommfont}
        \SetKwInput{KwInput}{Input}                
        \SetKwInput{KwOutput}{Output}              
        \DontPrintSemicolon
        \caption{Policy Addition // Policy Update} \label{alg:LayeredEncryption}
          \KwInput{$CT_{ABE_{1}}, APs, MPK$}
          \KwOutput{$nLayers, CT_{ABE_{2}}$}
          \For{$i \gets 0$ to $APs$}
          { 
              \If {$i == 0$}
              {
                $C \gets CT_{ABE_{1}}$ \;
                $u \gets H(C||AP_{i})$ \;
                $C_{i} \gets Enc_{ABE_{CP}}(MPK, AP_{i}, C, u)$ \;
                $nLayers++$\;
                \tcc*{nLayers reflects the current amount of encryption layers.}
              }    
              \Else
              {
                $u \gets H(C_{i-1}||AP_{i})$\;
                $C_{i} \gets Enc_{ABE_{CP}}(MPK, AP, C_{i-1}, u)$\;
                $nLayers++$\;
              }
            }
            $CT_{ABE_{2}} \gets C_{i}$\;
        \end{algorithm}

        \begin{enumerate}
            \item \replaced{The Internal CT Engine}{It} requests the layered $APs$ to the Policy Database.
            \item Then it performs the layered encryption over $CT_{ABE_{1}}$ to generate $CT_{ABE_{2}}$ by applying Algorithm \ref{alg:LayeredEncryption}. This same algorithm is the one used for Policy Update.
            \begin{enumerate}
                \item \added{The Internal CT Engine takes the layered $APs$.}
                \item \added{For the first iteration, it takes the inputed $CT_{ABE_{1}}$ and proceeds to rename it $C$.}
                \item \added{Then, for each layer $AP_{i}$, it creates the random nonce $u$ $\gets$ $H(C||AP_{i}).$ }
                \item \added{With $u$ generated, the internal CT Engine applies the CP-ABE encryption resulting in $C_{i}$. }
                \item \added{When $i = nLayers$, the iterations finish.}
                \item \added{Finally, Algorithm \ref{alg:LayeredEncryption} produces $CT_{ABE_{2}} \gets C_{i}$.}
            \end{enumerate}
             \item The Internal CT Engine returns $CT_{2} \gets (CT_{AES}, CT_{ABE_{2}})$.
             \item It finally sends $CT_{2}$ to the CT Database alongside information about the total number of layers, $nLayers$.
        \end{enumerate}

    \item $Decryption (CT, SK, nLayers) \rightarrow PT$ \par
    
            \begin{figure}[hbp]
            \centerline{\includegraphics[width=75mm]{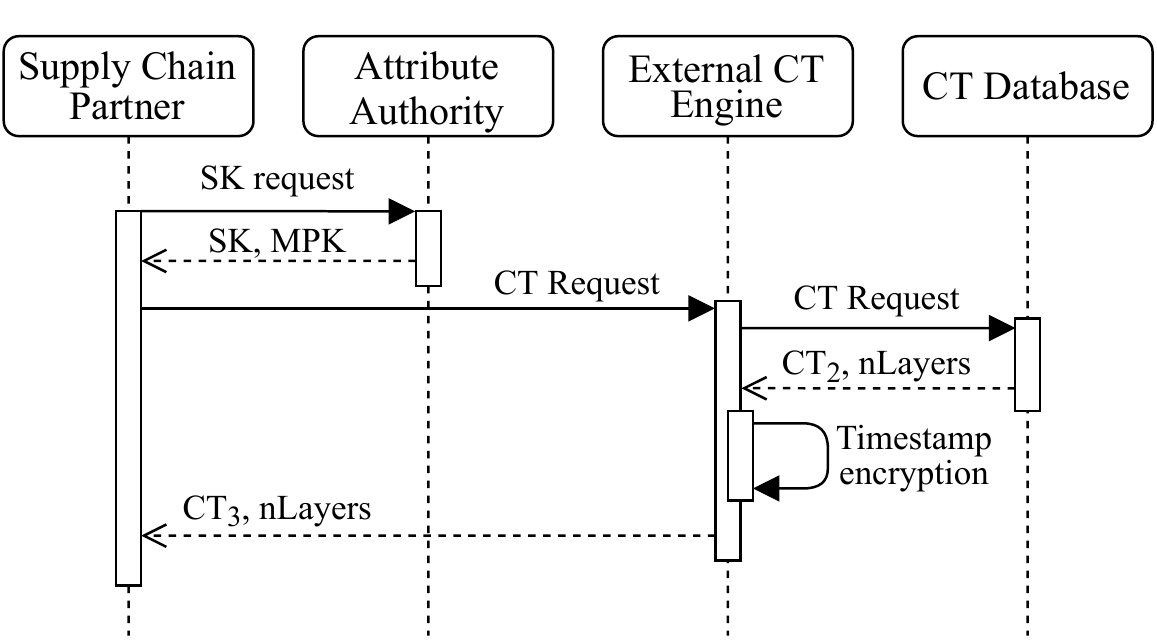}}
            \caption{Decryption message exchange}
            \label{fig:FIG6}
            \end{figure}
            In order to decrypt \added{the $SK_{sym}$ encapsulated in $CT_{2}$}, Supply chain consumers require a $SK$ provided by the AA. \replaced{Consumers require $SK$s to the AA whenever}{This action is performed whenever} they log for the first time into the system or after a system security incident, which requires $SKs$ to be reissued. \deleted{If they do not own one, they can request it from the AA.} 

            Once the consumers have the $SK$, they can request $CT_{2}$ and try to decrypt it. To ensure that the keys have been issued after the last security incident, the External CT Engine \replaced{performs a time-based encryption of $CT_2$. The result of this operation, $CT_{3}$, is sent back to the supply chain partner that requested it.}{performs the operations described in the following paragraphs.} The message exchange between the involved parties can be seen in Figure \ref{fig:FIG6} \added{and is explained below}.
            
            \begin{enumerate}
                \item The engine defines a new access policy that requires the $SK$ to have been generated after the last security incident. \replaced{For this purpose the policy takes the form of}{The defined policy is} $AP_{time} \gets (T_{SK} > T_{incident})$. \added{In it, $T_{incident}$ is the timestamp of the last registered security incident in the system. }
                \item \replaced{The External CT Engine}{It} generates the CP-ABE required random nonce computing $u \gets H(CT_{ABE_{2}}||AP_{time})$.
                \item The ciphertext is re-encrypted, \replaced{generating}{by applying} $CT_{ABE_{3}}$ $\gets$ $Enc_{ABE_{CP}}$ $(MPK, AP_{time}, CT_{ABE_{2}}, u)$.
                \item Finally, the engine sends the consumer the resulting ciphertext $CT_{3} \gets (CT_{AES}, CT_{ABE_{3}})$ and the \replaced{amount of policies contained in it ($nLayers$).}{updated layer number.}
            \end{enumerate}
            
            With the obtained $CT_{3}$, the consumer tries to decrypt it. \par
                    \begin{algorithm} [t]
        \SetKwInput{KwInput}{Input}                
        \SetKwInput{KwOutput}{Output}              
        \DontPrintSemicolon
        \caption{Layered Decryption // Policy Revocation} \label{alg:LayeredDecryption}
          \KwInput{$CT_{ABE_{3}}, nLayers, SK$}
          \KwOutput{$SK_{sym}', r', AAD, AP'$}
          $C \gets CT_{ABE_{3}}$ \;
          \For{$i \gets 0$ to $(nLayers-1)$}
          {
            $C_{i} \gets Dec_{ABE_{CP}}(MPK, SK, C_{i-1})$
           }
            $CT_{ABE_{1}} \gets C_{i}$\;
            $AAD \gets ExtractHeader(CT_{ABE_{1}})$ \;
            \deleted{$AP' \gets ExtracPolicy(CT_{ABE_{1}})$\;}
            $(SK_{sym}'||r') \gets Dec_{ABE_{CP}}(MPK, SK, CT_{ABE_{1}})$ \;
        \end{algorithm}

            \begin{enumerate}
                \item \replaced{The Consumer}{It} starts by applying Algorithm \ref{alg:LayeredDecryption} to $CT_{3}$, which outputs $SK_{sym}'$, $r'$, $AAD$, and $AP'$. This algorithm can be used for Policy Revocation of $i$ layers while $i < nLayers$.
                \begin{enumerate}
                    \item \added{The Supply Chain Partner takes $CT_{ABE_{3}}$ and loads it as $C_{i}$.}
                    \item \added{For every layer until $nLayers-1$, it decrypts $C_{i}$ by using its $SK$.}
                    \item \added {For the layer $nLayers -1$, the decryption of $C_{i}$ returns $CT_{ABE_{1}}$. $CT_{ABE_{1}}$ was the original ciphertext produced by the DO, and thus contains the $SK_{sym}$ used to generate the corresponding $CT_{AES}$.}
                    \item \added{The user extracts the header from $CT_{ABE_{1}}$ and stores it to use as $AAD$.}
                    \item \added{The partner performs the last CP-ABE decryption on $CT_{ABE_{1}}$, which returns $(SK_{sym}'||r')$.}
                    \item \added{Finally, Algorithm \ref{alg:LayeredDecryption} returns $AAD, SK_{sym}'$ and $r'$.}
                \end{enumerate}
                \item \replaced{The Supply Chain Partner performs a security encryption}{Following the ETSI~\cite{ETSI-CYBER2021AttributeV1.2.1} guideline, a security encryption is performed,} for which the nonce is defined as $u' \gets H(r'||AP'||SK_{sym}')$.
                \item The \added{security} encryption is computed as: $CT_{ABE_{1}}'$ $\gets$ $Enc_{ABE_{CP}}$ $(MPK, AP',$ $(r'||SK_{sym}'), u)$.
                \item If $ CT_{ABE_{1}}' ==  CT_{ABE_{1}}$, the consumer obtains $SK_{sym}$; if not, it obtains $\bot$.
                \item With \replaced{$SK_{sym}$}{the key}, the consumer can obtain the original message $PT \gets Dec_{AES}(SK_{sym}, CT_{AES}, AAD)$. \added{Note that each $PT$ is generated with a different $SK_{sym}$, so the decryption process has to be performed in its entirety for each encrypted data in the system.}

            \end{enumerate}

\end{itemize}

%% file: Docs/6.Experimental.tex
\section{Experimental evaluation}\label{Sec:6}
To verify the requirement fulfillment analyzed in Section \ref{Sec:4.2}, we have created a setup simulating the roles of the proposed system. \added{Subsection \ref{Sec:6.1} describes the testbed setup. Section \ref{Sec:6.2} discusses the designed experiments and the measured parameters.}

\added{Regarding the requirements established in section \ref{Sec:3.1}}, \replaced{\textbf{``R1. Heterogeneous Data Exchange"}}{\textbf{R1}} is fulfilled by the message exchange system since it \replaced{considers the three data types identified in Section \ref{Sec:3.1}, i.e., production parameters, digital products, and data related to physical product exchange.}{is designed taking into account the three types of identified data.} Meanwhile, the combination of CP-ABE and AES-GCM \replaced{tackles \textbf{``R3. E2E data confidentiality and integrity".}}{provides \textbf{R3}.} Moreover, CP-ABE also fulfills \replaced{\textbf{``R4. Non-Identity Based Access To Data"}}{\textbf{R4}} by creating an ABAC~\cite{waters2011ciphertext}.

\replaced{Thus, this section evaluates the fulfillment of \textbf{``R2. Computational Scalability"} and \textbf{``R5. System flexibility and responsiveness"}.}{In these experiments, we focus on the fulfillment of\textbf{R2.} and \textbf{R5}.} 

\subsection{Testbed Setup} \label{Sec:6.1}
The main elements of the testbed designed for the experimental evaluation are:
\begin{itemize}
    \item \textbf{DO}: The tasks assigned to this role are \replaced{performed}{accomplished} by Raspberry Pi Zero W (RPI0-W) with Raspbian Stretch. It has 512MB RAM, 1GHz, a single-core ARMv6, and Wi-Fi. It is thus a good representation of an IIoT device. The RPI0-W applies AES-GCM to messages and the first layer of CP-ABE Encryption to the $SK_{sym}$. 
    \item \textbf{Internal CT Engine}: A Raspberry Pi 4 (RPI4) with 32 bits Ubuntu Server TLS. The RPI4 has a 8GB LPDDR4-3200 SDRAM and a Quad core Cortex-A72 (ARMv8). The RPI4 has to add the policy layers to the encrypted $SK_{sym}$.
    \item \textbf{Consumer}: This role is simulated using an Ubuntu Virtual Machine with an allocated RAM of 4GB within a host machine with an Intel Core i7-8850H CPU processor. 
\end{itemize}

The used library is a modified version of OpenABE~\cite{Zeutro2018} that implements the Multi-Layered CP-ABE. Further modifications have also been applied to compile the library for Raspbian, as well as ARMv6 and ARMv8 architectures.

\subsection{Experiments definition} \label{Sec:6.2}

\replaced{Regarding \textbf{R5}, its fulfillment is covered by the use of Multi-Layered CP-ABE, with the only constraint on its fulfillment being the system's capability to accomplish the assigned tasks efficiently. Meanwhile, the system must perform encryption efficiently to be considered \textbf{R2} compliant.}{To be considered \textbf{R2} compliant, the system must perform encryption efficiently. Similarly, if \textbf{R2} is achieved, we consider \textbf{R5} fulfilled as well. This requirement is covered by the use of Multi-Layered CP-ABE, with the only constraint on its fulfillment being the system's capability to accomplish the assigned tasks efficiently.}

\replaced{System DOs must encrypt two items. The original message, known as Plaintext ($PT$), and the symmetric key, $SK_{sym}$. The $PT$ encryption time is directly related to the size of the $PT$. Something similar happens with $SK_{sym}$ encryption, which also depends on its length. Therefore to study \textbf{R2} compliance, the results focus on the time required for $SK_{sym}$ encryption. This is because the PT size is highly variable, while the chosen $SK_{sym}$ is always 256 bits.}{Since our system uses a hybrid encryption format, the DO encrypts two objects: the message $PT$ and the symmetric key $SK_{sym}$. The time to encrypt $PT$ varies depending on its size, and the time required to encrypt the AES key does not correlate with the size of $M$. Hence, measuring the time for $M$ encryption generates results not transferable to other systems. Therefore, to verify the \textbf{R2} compliance, only the $SK_{sym}$ encryption is analyzed.}

\replaced{We measure the encryption time required by DO (RPI0-W) to generate $CT_{ABE_{1}}$, and by the internal CT engine (RPI4) to generate $CT_{ABE_{2}}$. In addition, the result of this measurement is compared with the time required to create $CT_{ABE_{2}}$ in the DO. The comparison illustrates the amount of work that has been offloaded to the internal CT engine.}{We measure the encryption time for the DO (RPI0-W) to generate $CT_{ABE_{1}}$ and for the Internal CT Engine (RPI4) to generate $CT_{ABE_{2}}$. The encryption is not considered complete until $CT_{ABE_{2}}$ has been obtained since the $CT$ includes the updatable policies. In addition, the result of this measurement is compared to generating $CT_{ABE_{2}}$ in the DO. The comparison shows how much work has been offloaded to the Internal CT Engine.}

\replaced{To analyze how much the layers affect the Internal CT Engine, we measure the time required for the Internal CT Engine to generate $CT_{ABE_{2}}$ if it were to apply all attributes in a single layer. In addition, the time needed to add policies represents the time it takes to update $CT_{ABE}$ since the update is performed by the same device using the same algorithm.}{Similarly, we also measure the time required by the Internal CT Engine to generate $CT_{ABE_{2}}$ by applying all the attributes in one layer. This analysis shows how much the Internal CT Engine is affected by the application of the layers. The time required to add policies is considered representative of the time it takes to update the $CT_{ABE}$ since the update is performed by the same device using the same function.}

\replaced{On the other hand, as already mentioned, $SK_{sym}$ is encrypted using CP-ABE. CP-ABE expands the original message when generating the $CT$, so this expansion has been studied as a function of the number of layers added. Furthermore, it has been compared with the result of encrypting a 160kBytes message to understand to what extent this growth of $CT_ABE$ affects the $CT = (CT_{AES}, CT_{ABE})$ obtained by consumers.}{On the other hand, as already mentioned, $SK_{sym}$ is encrypted using an asymmetric encryption algorithm. Asymmetric algorithms generate a $CT$ larger than the original message they encrypt. This expansion of $CT_{ABE_{2}}$ has also been studied in terms of the number of layers added. In addition, it has been compared with the result of encrypting a 160kBytes message to understand to what extent this growth affects the $CT = (CT_{AES}, CT_{ABE})$ that consumers obtain.}

\added{The measurements explained above have been obtained by performing the layered encryption 500 times and obtaining the average operation time. In addition, to correctly see the evolution of the results, we added up to 15 layers formed by three attributes. Thus, for the worst case, a total of 45 attributes are used.}

\subsection{Discussion of the Results} \label{Sec:6.3}

One of the main requirements of industrial scenarios is the availability of the system. Therefore, to fulfill \textbf{R2}, the IIoT devices have to encrypt information \replaced{with an acceptable delay.}{without delay.} 

\replaced{As mentioned, the proposed system combines AES-GCM and CP-ABE. Every encryption system causes an expansion of the message it encrypts. That is, the $CT$ will always be larger than the corresponding $PT$. However, the expansion generated by AES on the original message is practically negligible. In contrast, the expansion generated by CP-ABE cannot be disregarded. Therefore, we study how this increase in the $CT$ size affects IIoT devices.}{As mentioned, the proposed system combines a symmetric and asymmetric encryption algorithm. Applying symmetric encryption to a message results in a $CT$ with less overhead than the one created by the asymmetric algorithm. Therefore, we study how the IIoT devices are affected by the application of CP-ABE rather than AES-GCM.}

\begin{figure}[!htbp]
\centerline{\includegraphics[width=90mm]{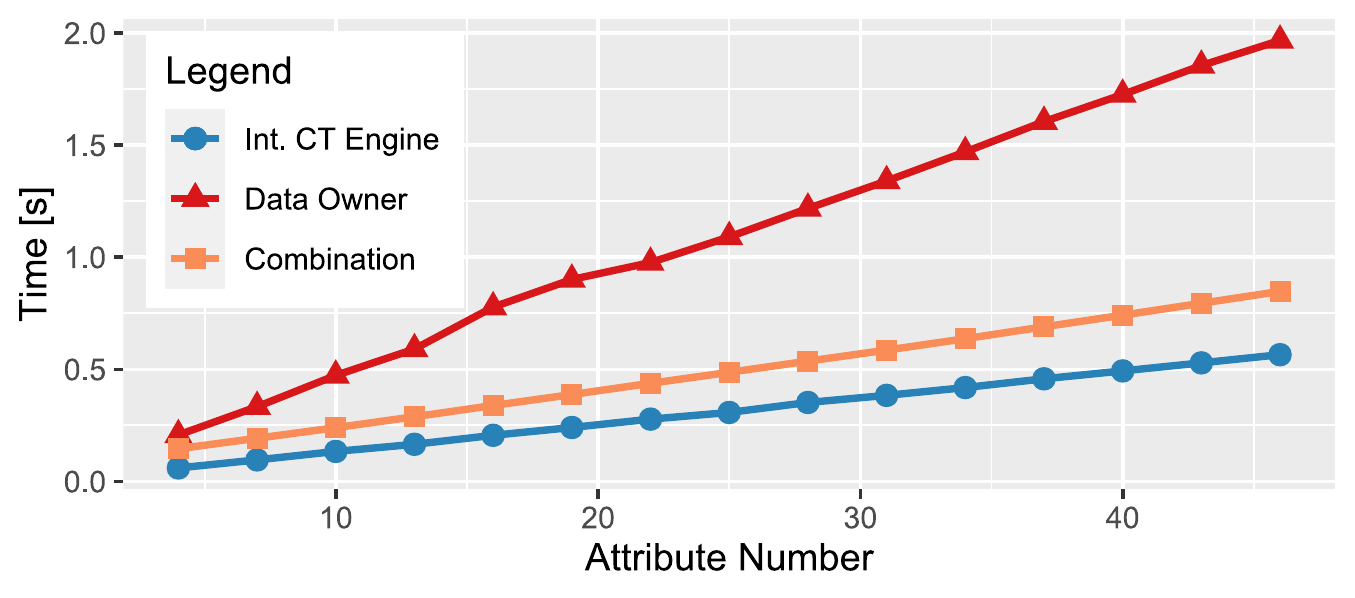}}
\caption{Total time required to generate $CT_{ABE_{1}}$ and transform it to $CT_{ABE_{2}}$ in three scenarios: One layer in DO (red triangle), one layer in Int. CT Engine (blue dot) and the combined use of DO and Int. CT Engine (orange square).}
\label{fig:FIG7}
\end{figure}

\replaced{Figure \ref{fig:FIG7} shows the time required to encrypt $SK_{sym}$ on the y-axis, while the x-axis represents the total number of attributes contained in the policy. As mentioned above, the original message is encrypted with AES, so the layered encryption only affects the $SK_{sym}$ encryption times. This proposal uses AES-256-GCM, so $SK_{sym}$ is always 256 bits, regardless of the message size that AES encrypts. Therefore, Figure \ref{fig:FIG7} only shows the time required for $SK_{sym}$ encryption since it is the only value affected by the layered encryption. The encryption times in Figure \ref{fig:FIG7} should be added to those resulting from symmetric encryption.}{Figure \ref{fig:FIG7} shows the required time to protect $SK_{sym}$ versus the attributes contained in the policy. As mentioned above, the original message size has a substantial impact on encryption times, and thus it is considered more relevant to show the $SK_{sym}$ encryption times. This proposal uses AES-256-GCM, so $SK_{sym}$ is 256 bits long. Therefore, the times shown in Figure \ref{fig:FIG7} are added in every instance to the message encryption, regardless of the original message size.} 

Figure \ref{fig:FIG7} shows \added{with red triangles} the result of the DO applying the entire policy $AP=(Att_{1}$ \textit{AND} $Att_{2}$ \textit{AND ... AND} $Att_{n})$. Due to its limited capabilities, \deleted{although the IIoT device can encrypt,} \replaced{for 10 attributes, it takes 100\% more time compared to the combined use case (in orange squares). In fact, for 40 attributes the difference goes from 750ms to 1.75s, increasing a 133\%. }{it takes almost half a second.} \deleted{Because of this, we do not consider \textbf{R2} met, as these times are not suitable for use in an industrial environment.}

\replaced{In the combined used case, the DO applies the first policy $AP_{1}= (Att_{1}$ \textit{AND} $Att_{2}$ \textit{AND} $Att_{3})$.}{The same chart also presents the combination of the DO and the Internal CT Engine to solve this. In this case, as explained in Section \ref{Sec:5}, the DO applies the first policy $AP_{1}= (Att_{1}$ \textit{AND} $Att_{2}$ \textit{AND} $Att_{3})$.} Then the Internal CT Engine applies the next layers by setting 3-attribute policies: $AP_{2} = (Att_{4}$ \textit{AND} $Att_{5}$ \textit{AND} $Att_{6})$, $AP_{3}= (Att_{7}$ \textit{AND} $Att_{8}$ \textit{AND} $Att_{9})$, up to $APn$. \deleted{For this case, the x-axis represents the total number of attributes. That is, in the case of $AP_{1}$, $AP_{2}$ and $AP_{3}$, the corresponding point on the horizontal axis would be \textit{Attribute Number = 9}.} The combined use of DO and Internal CT Engine demonstrates that even the worst-case encryption \added{(45 attributes)} takes 56\% less time compared to the sole use of DO: 2s for the IIoT device and 0.84s for the combined case. It is also observed that the Internal CT Engine does not take much longer to add several layers \replaced{(represented with orange squares)}{(orange case)} than it does to apply a single layer \replaced{(represented with blue dots)}{(blue case)}. It is also noteworthy that the DO and Internal CT Engine combination achieves full key encryption in less than one second, even in the worst case.

\begin{figure}[!htbp]
\centerline{\includegraphics[width=90mm]{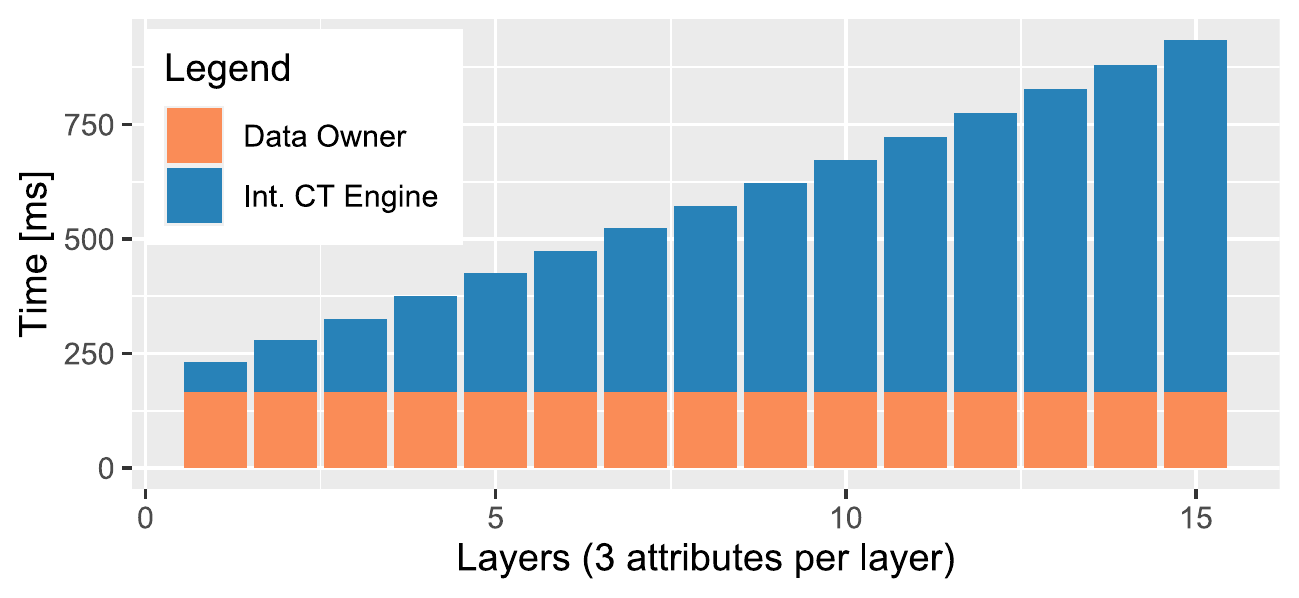}}
\caption{Total time required by the DO to generate $CT_{ABE_{1}}$ (orange) and the total time required for the Int. CT Engine to generate $CT_{ABE_{2}}$ (blue).}
\label{fig:FIG8}
\end{figure}

To better understand the combination of DO and External CT Engine and its correlation with encryption times, we present Figure \ref{fig:FIG8}. It provides a better understanding of the scenario \replaced{represented with orange squares}{presented in orange} in Figure \ref{fig:FIG7}. DO protects the $SK_{sym}$ with a policy $AP_{1}= (Att_{1}$ \textit{AND} $Att_{2}$ \textit{AND} $Att_{3})$. Then, the Internal CT Engine adds layer-by-layer policies that maintain the same format. Figure \ref{fig:FIG8} shows the total time required to create $CT_{ABE_{1}}$ and generate $CT_{ABE_{2}}$. This Figure depicts how time is distributed between the DO and the Internal CT Engine. The time consumed by the IIoT device in this scenario is constant, while the time consumed by the Internal CT Engine grows linearly as layers are added. This linear growth implies that the required time to add more layers is predictable and that encryption times do not escalate out of control.

\begin{figure}[!htbp]
\centerline{\includegraphics[width=90mm]{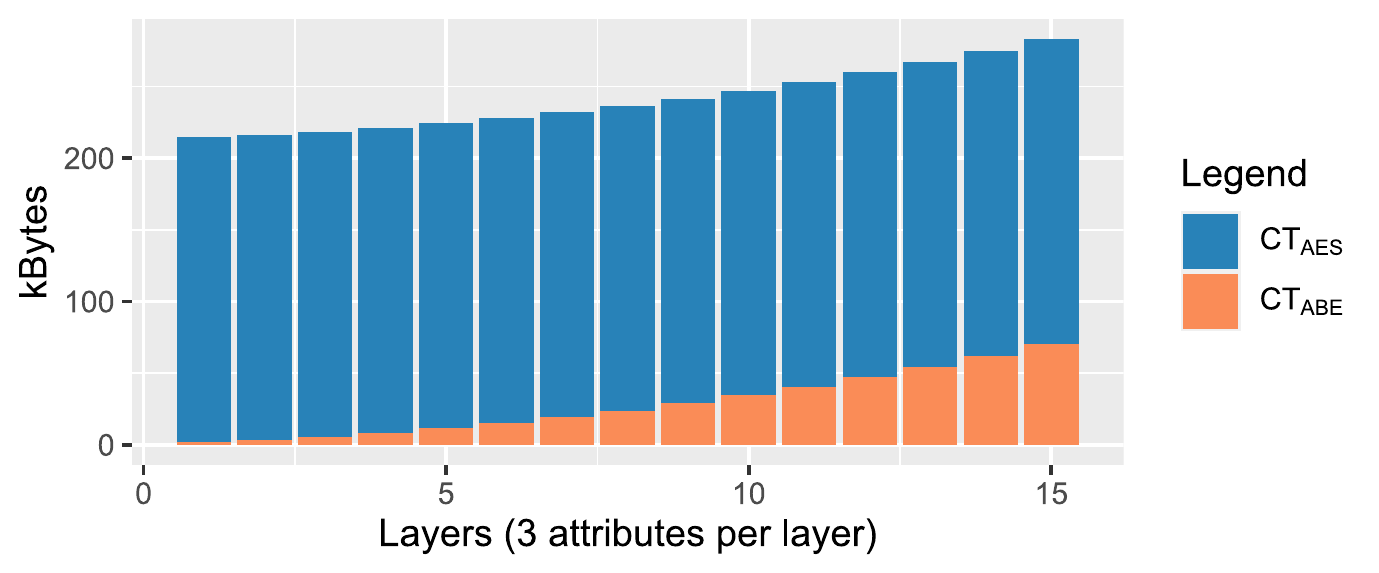}}
\caption{Sizes of the generated $CT_{ABE}$ and $CT_{AES}$}
\label{fig:FIG9}
\end{figure}

\replaced{As discussed in Section \ref{Sec:6.2}, we analyze the size of $CT_{ABE_{2}}$. Figure \ref{fig:FIG9} shows the total length of a $CT$ requested by a consumer for an original message of 160 kBytes. The figure compares the size in kBytes of the $CT$ with the total number of layers contained in $CT_{ABE_{2}}$.}{As already discussed in \ref{Sec:6.2}, we analyze the size of $CT_{ABE_{2}}$. Figure \ref{fig:FIG9} shows the total size of the $CT$ to be requested by the Consumer for an original message of 160 kBytes. The size in kBytes is shown as a function of the total number of layers contained in  $CT_{ABE_{2}}$.} In the same figure, it can also be seen how the message size is divided between $CT_{AES}$ and $CT_{ABE}$. $CT_{ABE}$ has a significantly larger size than the original 256bits, but the expansion loses relevance compared to the size of the final $CT$. The reason for this is that $CT$ is influenced by the much larger size of $CT_{AES}$. \replaced{Figure \ref{fig:FIG9} shows that the proposed system achieves its maximum efficiency for large data packets, where the original message is much larger than the 256-bit AES key. Therefore, this result demonstrates the proposal's suitability for the industrial environment. The DOs could send larger packets in longer intervals, which would reduce the interaction with the IIoT devices that constitute the DOs.}{These results show that the proposed system achieves its highest efficiency for large data packets. This combines well with the proposed system since the DOs can send larger packets in longer time intervals, reducing the interaction with DOs.} In addition, the algorithm used to add policies is the same as the one used to update them, so it is proven that the update does not generate a $CT_{ABE}$ larger than $CT_{AES}$ \added{for a reasonable number of attributes}.




%% file: Docs/7.Conclusions.tex
\section{Conclusions} \label{Sec:7}
This paper proposes an E2E secure data exchange system for a value chain. E2E security protects members of the value chain from compromised partners and reduces the amount of sensitive data exposed due to data breaches. The proposed exchange system provides DOs in the value chain with control over who can access the information while allowing a straightforward definition of new value-chain partners. Having scalability to add new users favors system efficiency, flexibility, and management. The proposal is also mapped to ETSI's High-Level Reference Model, facilitating the integration of the proposed system in an industrial environment

The system is designed according to five requirements defined after studying the literature and considering security and industrial requirements. Our system fulfills \replaced{\textbf{``R1. Heterogeneous Data Exchange"}}{\textbf{``R1. Product and data exchange"}} because, as shown in the Results, even IIoT devices can perform the tasks assigned to DOs. \replaced{IIoT devices are the ones that generate production parameters and data related to the exchange of physical products. This type of device has limited resources. Therefore, if they can protect the data they generate, the exchange of digital products is also protected since the devices that manage the latter have a higher computational capacity than IIoT devices.}{IIoT devices generate production parameters and data related to physical product exchange. Digital product exchange is also protected since IT devices have more significant computing capabilities than IIoT devices.} 

Similarly, experiments also show the system is \replaced{compliant with \textbf{``R2. Computational Scalability"}}{\textbf{``R2. Working with limited computational power"}} compliant. The IIoT devices can perform the designed tasks, and its combined use with a more powerful device yields efficient results. At any rate, the experiments also show that IIoT devices can protect the $SK_{AES}$, attaining E2E security. It also means that \textbf{``R3. E2E data confidentiality and integrity"} is fulfilled since only the DO and the consumer know $SK_{AES}$. Integrity is achieved by employing AES-GCM as symmetric cipher.

The system uses Multi-Layered CP-ABE to protect $SK_{AES}$, fulfilling \textbf{``R4. Non-Identity Based Access Control to Data"} by having a RBAC to data and \textbf{``R5. Access Policy Update"} by achieving policy update without breaking E2E Security. Besides, the IIoT devices do not intervene in the policy update, which reinforces the compliance with \textbf{R2}.

Finally, the system also denies users using old $SK_{ABE}$ by generating an $AP$ requiring a $SK_{ABE}$ generated after the last security event in the system.

%% file: Docs/ACKs.tex
\section*{Acknowledgment}
The European commission financially supported this work through Horizon Europe program under the IDUNN project (grant agreement N$^{\circ}$ 101021911). It was also partially supported by the \textit{Ayudas Cervera para Centros Tecnológicos} grant of the Spanish Centre for the Development of Industrial Technology (CDTI) under the project EGIDA (CER-20191012), and by the Basque Country Government under the ELKARTEK program, project REMEDY - REal tiME control and embeddeD securitY (KK-2021/00091).